\documentclass[aip,apl,reprint,groupedaddress]{revtex4-1}

\usepackage{graphicx}

\begin{document}

%\preprint{}

\title{A microfabricated ion trap with integrated microwave circuitry}

\author{D. T. C. Allcock}
\author{T. P. Harty}
\author{C. J. Ballance}
\author{B. C. Keitch}
\altaffiliation{Present address: Department of Physics, ETH Z\"{u}rich, Switzerland}
\author{N. M. Linke}
\author{D. N. Stacey}
\author{D. M. Lucas}
\email{d.lucas@physics.ox.ac.uk}

\affiliation{Department of Physics, University of Oxford, Clarendon Laboratory, Parks Road, Oxford, OX1 3PU, United Kingdom}

\date{\today}

\begin{abstract}

We describe the design, fabrication and testing of a surface-electrode ion trap, which incorporates microwave waveguides, resonators and coupling elements for the manipulation of trapped ion qubits using near-field microwaves.  The trap is optimised to give a large microwave field gradient to allow state-dependent manipulation of the ions' motional degrees of freedom, the key to multiqubit entanglement.  The microwave field near the centre of the trap is characterised by driving hyperfine transitions in a single laser-cooled $^{43}$Ca$^+$ ion.  

\end{abstract}

\maketitle

Manipulation of the internal and external degrees of freedom of trapped atoms with electromagnetic fields lies at the heart of many modern atomic physics experiments\cite{Wineland11}, with applications in precision spectroscopy and metrology, quantum information processing, and the study of degenerate quantum gases. Such manipulation is usually achieved using laser light and/or free space microwave or radiofrequency radiation. With the advent of planar ion\cite{Sei06} and neutral atom chip traps\cite{AtomChips}, which confine atoms tens of microns above a surface, it has also become possible to use near-field radiation above microwave conductors.  This near-field microwave regime has three qualitative differences to the free space regime.  Firstly, the magnetic fields that can be created with a given power are orders of magnitude higher, allowing for high Rabi frequencies\cite{Osp11}.  Secondly, much larger field gradients can be produced than in a free space microwave field because the spatial variation of the field is determined by the dimensions of the conductors rather than the wavelength of the radiation. These gradients can be used to produce state-dependent forces with application to quantum logic gates\cite{Osp11} or atom interferometry\cite{Boh09}, for example. Thirdly, the localisation of the field close to the conductor allows spatially selective addressing of atoms in different regions of the chip, essential for scalable quantum computing. Although state-dependent forces and selective addressing are routinely produced by laser beams, control of the optical frequency, phase and intensity at the levels required for fault-tolerant quantum computing are much more challenging than in the microwave regime.

%\section{Trap design}

\begin{figure}
\includegraphics[width=\columnwidth]{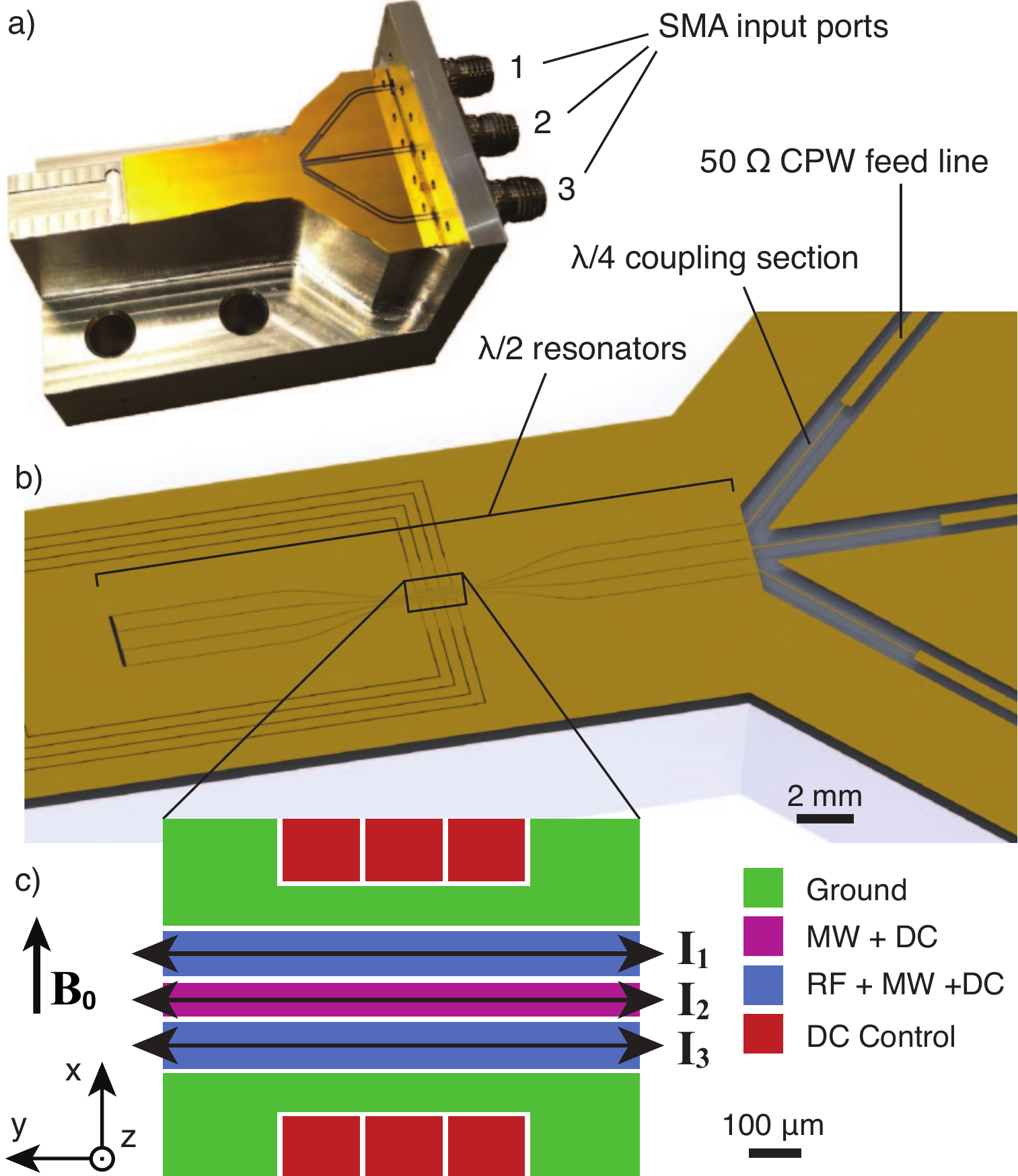}
\caption{\label{trap}a) The trap substrate mounted on its aluminium base. b) A schematic showing the resonators, the $\lambda/4$ coupling elements and the coplanar waveguides.  c) The central trapping region showing the microwave currents, I$_1$, I$_2$, I$_3$, and the static magnetic field direction, \textbf{B}$_0$.  Ions are trapped 75\,$\mu$m above the plane of the electrodes.} 
\end{figure}

The near-field regime was first explored using neutral atom chips with integrated coplanar waveguides (CPW)\cite{Boh09}, and subsequently in surface-electrode ion traps, either by driving microwave currents in simple electrode structures\cite{Osp11, Bro11} or by incorporating a CPW into the trap structure\cite{Ant11}. In the trap described here we take near-field techniques a step further and integrate more complex microwave circuitry, including a set of three resonators, in a surface-electrode ion trap.  The frequency of the resonators was designed to be 3.22\,GHz to match the hyperfine splitting between the F=3 and F=4 manifolds in the $4S_{1/2}$ ground level of $^{43}$Ca$^+$.  Transitions between these two manifolds can be utilised as a very robust qubit with coherence times of over a second reported\cite{Lucas2007}.  The resonators allow us to design the trap such that, on resonance, it is a $50\,\Omega$ load, impedance-matched to the microwave drive circuitry. This improves stability by reducing unwanted microwave reflections from the trap.  The build-up in the resonator also reduces the microwave power required to achieve a given current.  Suppressing unwanted carrier transition when driving motional sideband transitions\cite{Osp08} is a crucial feature in several quantum logic gate implementations.  In order to achieve this we use three resonators side-by-side (see figure~\ref{trap}) so that, by tuning the relative phase and amplitudes, we can null the microwave field at the position of the ion whilst retaining a large microwave field gradient.

%\subsubsection{Layout of resonators}

\begin{figure}
\includegraphics[width=\columnwidth]{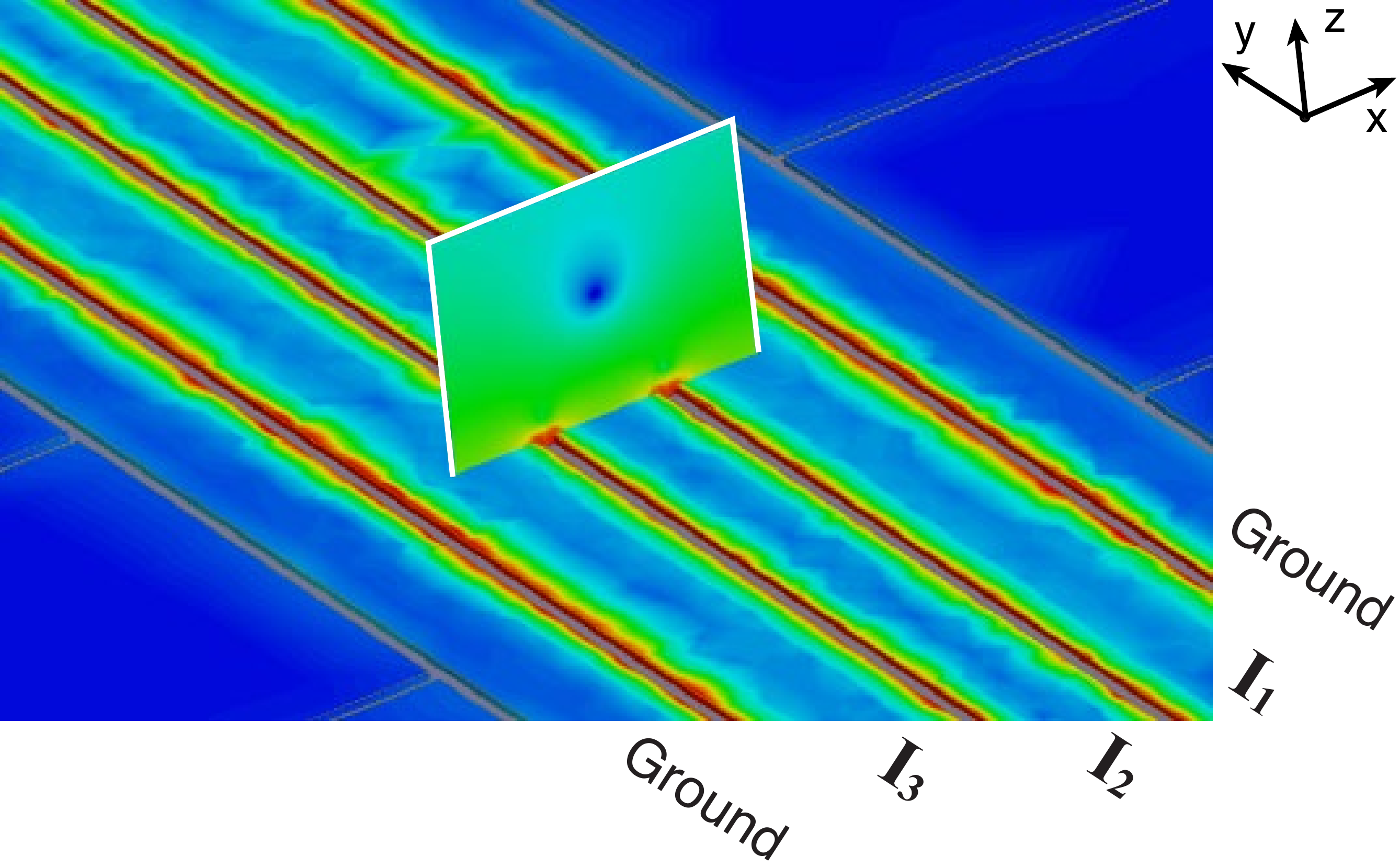}
\caption{\label{fields}HFSS simulation of the microwave surface current density on the trap electrodes and the magnitude of the microwave magnetic field in the radial plane of the ion (rectangular inset in figure) when the field is nulled at the ion.  Note the crowding of the current close to the edges of the electrodes and the current flowing in the ground plane.}
\end{figure}

The layout of the microwave resonators was optimized to produce the largest microwave field gradient at the ion for a given microwave current density (the likely limiting factor in how much microwave power can be applied to the trap).  This optimization required finite element simulations (using Ansys HFSS software) because at microwave frequencies the skin effect and proximity effect produce a complex distribution of currents in each conductor as well as induced currents in other nearby conductors (see fig.~\ref{fields}) that cannot be calculated analytically.  As we require two radio-frequency (RF) electrodes in a fixed position in order to produce the required trapping pseudopotential, there are a limited number of choices for the layout of the three microwave resonators.  Combining the resonators with the RF electrodes rather than placing separate microwave electrodes around the RF electrodes was shown in simulations to give a larger microwave field gradient at the ion.  The application of an RF potential then prevents us from grounding the resonators at any point.  Thus the smallest resonant structure we can use is an open-ended half-wave resonator, with the ions in the centre (directly above the current anti-node).  Microwave power is brought to the resonators on three $50\,\Omega$ CPW feed lines.  These cannot be capacitively coupled to the resonators from one end\cite{Poz05} as this would preclude application of DC and RF trapping potentials to the electrodes.  Instead we use a $\lambda/4$ section of CPW of higher impedance, which is electrically equivalent to a capacitor\cite{Zhu05} (see fig.~\ref{trap}b).

Figure~\ref{trap} shows the final design in which critical dimensions, such as the length of the resonators and coupling elements, were optimised to give the largest microwave field gradient at the ion.  The RF null, where the ions are trapped, was chosen to be 75\,$\mu$m above the surface.  The trap has 91\,$\mu$m wide RF electrodes, a 70\,$\mu$m wide centre electrode and 10\,$\mu$m gaps between all electrodes.  As well as the basic elements discussed above, there are several other notable design features.  The thickness of the electrodes was chosen to be 5\,$\mu$m because the skin depth at 3.22\,GHz is 1.4\,$\mu$m; thicker electrodes would not significantly reduce losses but would be harder to fabricate.  The DC electrodes are set back from the RF electrodes with a 75\,$\mu$m wide section of ground plane in between, since otherwise the gaps between DC electrodes would prevent current flowing in the ground plane (see fig.~\ref{fields}), causing a reflection of microwave power.  The microwave electrodes are also flared out away from the trap region to reduce resistive losses and spread out the thermal load.

%\section{Implementation}

%\subsection{Trap fabrication}

The electrodes are electroplated gold (fabricated as in previous work\cite{Allcock10}) on a 0.5\,mm thick c-cut sapphire substrate.  Sapphire has low microwave loss, good thermal conductivity and its high dielectric constant reduces the length of the resonators.  The trap is epoxied to an aluminium mounting block which incorporates SMA connectors that are glued directly to the trap with conductive epoxy.  These are then connected to a UHV feedthrough with vacuum-compatible coaxial cables (combined cable and feedthrough insertion loss is $<0.6$\,dBm at 3.22\,GHz).  

%\subsection{RF/microwave diplexer}

\begin{figure}
\includegraphics[width=\columnwidth]{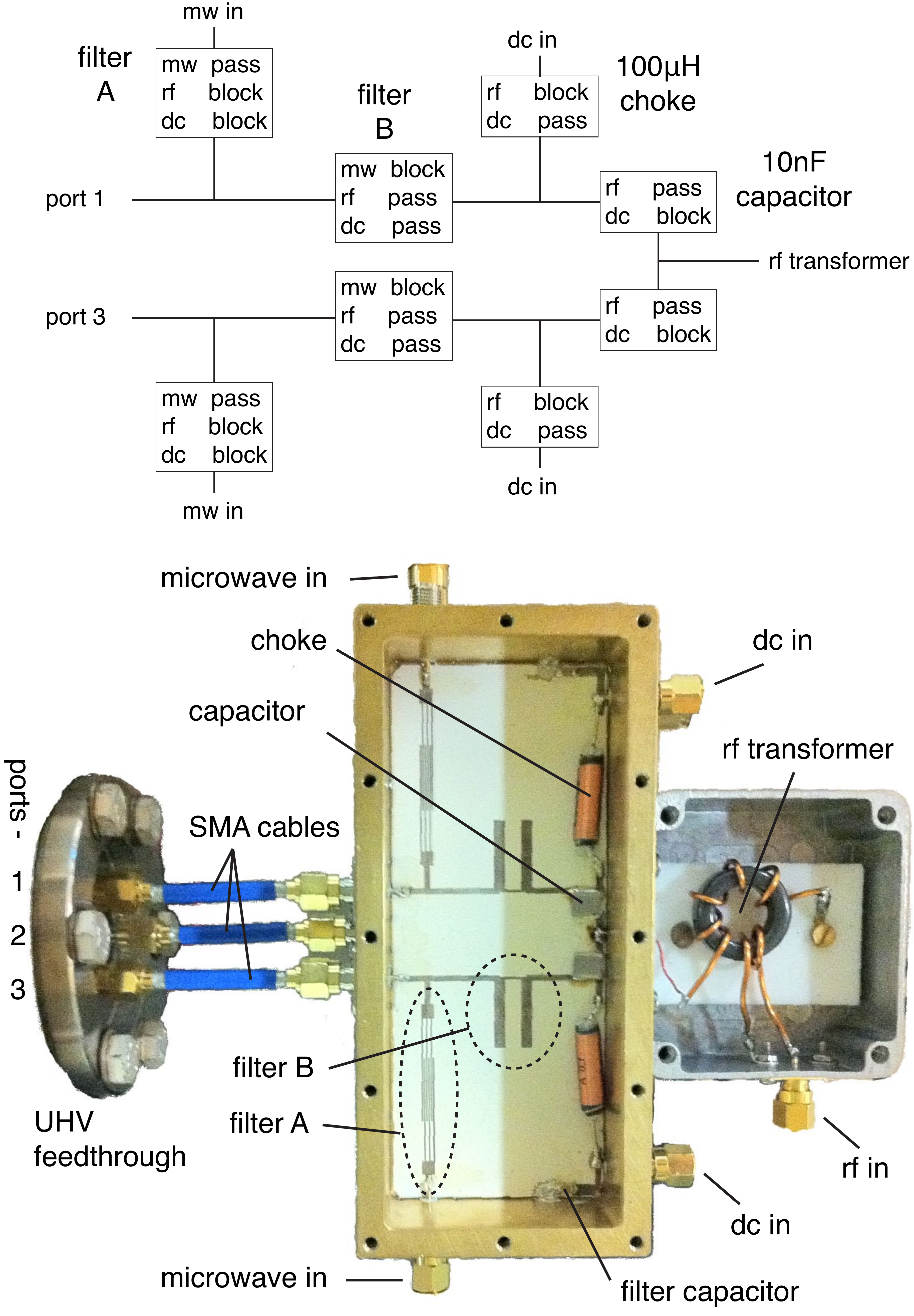}
\caption{\label{diplexer}Schematic (top) and photograph with metal cover removed (bottom) of diplexer.  The third diplexer channel for the centre trap electrode (port 2) is mounted in a separate box below the two channels shown.  It is identical except that the RF connection is grounded.}
\end{figure}

The design of the trap requires each of the three central trap electrodes to be connected to the microwave drive circuitry, a DC bias voltage and the trapping RF voltage (or an RF ground path in the case of the centre electrode).  This is done via an external diplexer circuit made up of two microstrip filters (see fig.\,\ref{diplexer}).  Filter A is a coupled line bandpass filter which passes 3.22\,GHz but is open circuit for RF and DC.  Filter B is a microstrip stub bandstop filter at 3.22\,GHz but passes RF and DC.  The microwave insertion loss of the diplexer is measured to be $<$0.9\,dB at 3.22\,GHz.  The RF is connected via 10nF capacitors to allow a separate DC bias to be added to each electrode via a 100\,$\mu$H choke. The diplexer circuit was designed using Microwave Office software and is fabricated on Rodgers RO3003 circuit board.  A transformer is used to impedance match the 50\,$\Omega$ RF amplifier to the trap and step-up the voltage.  This consists of an iron powder toroid (Micrometals T80-10) in a resonant circuit configuration with a $Q$ of 65 and a voltage step-up of 9.6 at 38.7\,MHz.  The trap DC electrodes are grounded at RF and microwave frequencies by wirebonding them to 1nF single-layer capacitors glued directly to the trap mount.  Further details on the testing of the diplexer, trap electrical connections and the vacuum system are described elsewhere\cite{Mine}.

%\section{Testing}

%\subsection{Electrical testing}

\begin{figure}
\includegraphics[width=\columnwidth]{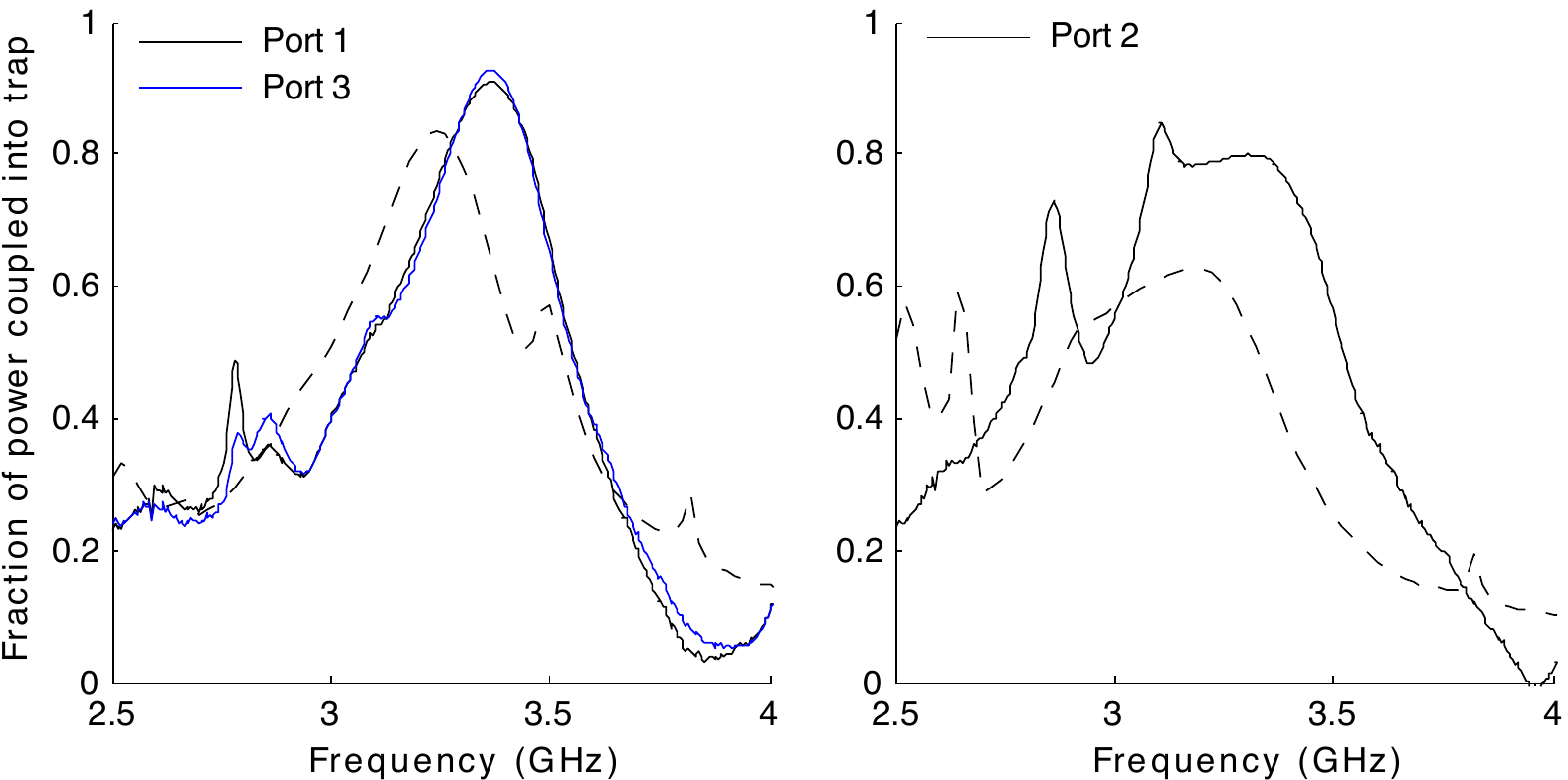}
\caption{\label{power}Comparison of VNA data (solid line) and the HFSS simulation (dashed line) of the total fraction of power incident on a given microwave port that is coupled into the trap.}
\end{figure}

A vector network analyser (VNA) was used to measure the fraction of microwave power coupled into the trap on each port.  This is compared with the simulation in fig.\,\ref{power}.  The $Q$ of the microwave resonators is approximately 6, matching the simulation well.  The resonant frequency of the resonators is around 5\% higher than simulated.  We believe this is due to inadequate refinement of the simulation rather than dimensional inaccuracies in the trap fabrication\cite{Mine} and could be corrected for in a second version of the trap.  In any case, the modest $Q$ of the resonators means the required frequency is close enough to resonance that we couple most of the incident power into the resonators ($>68\%$ at 3.22\,GHz). 

%\subsection{Ion loading and lifetime}

$^{43}$Ca$^+$ ions are loaded using isotope-selective photoionisation\cite{Lucas04} from an isotopically-enriched source (12\% $^{43}$Ca).  At an RF frequency of 38.7\,MHz and voltage amplitude of 72\,V, the radial secular frequency is 4\,MHz, the stability parameter is $q$=0.3 and the trap depth is 59\,meV.  Axial secular frequencies are up to 500\,kHz with DC control voltages of $<10\,$V.  The separate DC bias available on each RF electrode allows a DC quadrupole to be applied in order to rotate and split the frequency of the two radial modes\cite{Allcock10}.  The cooled single-ion lifetime is several hours at a pressure of $<10^{-11}\,$Torr.

%\subsection{Heating rates}

Using a modified\cite{Allcock10} Doppler-recool technique\cite{Wes07} we measure a heating rate  of $\dot{\overline{n}}=1.4(3)$\,quanta/ms on the axial mode at a secular frequency of $\omega = 2\pi\times500\,$kHz. This corresponds to an electric field spectral noise density of $\omega.S_E(\omega)= 1.6(3)\times10^{-5}\,$V$^2/$m$^2$, which is extremely low for a room-temperature trap of this ion-electrode separation and comparable with results from cryogenically-cooled traps (see reference\cite{Hite12} for a recent survey).

%\subsection{Micromotion}

The micromotion compensation indicates that the electric field offset at the ion is stable to better than $\pm$5\,V/m ($\lesssim30$\,nm displacement from RF null) in the radial direction parallel to the trap substrate over several weeks, with no detectable change after loading an ion.  A differential phase shift between the two RF electrodes, which would cause uncompensatable micromotion \cite{Ber98}, was avoided by keeping both RF paths through the diplexer as similar as possible.  Experimentally the uncompensatable micromotion is close to the limit of our detection sensitivity ($\sim2$\,nm amplitude).

%\subsection{Microwave field measurements}

Measurements of the microwave field amplitude were made by observing Rabi flopping between ground level hyperfine states in a single $^{43}$Ca$^+$ ion at a static field of \textbf{B}$_0$=2.3\,mT.  By measuring the Rabi frequencies of both a $\pi$ ($S^{4, 0}_{1/2}\leftrightarrow S^{3, 0}_{1/2}$, where the superscript indicates the angular momentum quantum numbers $F$, $M_F$) and a $\sigma$ ($S^{4, 0}_{1/2}\leftrightarrow S^{3, +1}_{1/2}$) microwave transition we were able to calculate the projection of $\mathbf{B}_{\mu w}$ perpendicular and parallel to \textbf{B}$_0$.  These two transitions are separated by 8.1\,MHz (3.226 and 3.218\,GHz respectively) but the frequency dependence of the resonators is small over this range.  Comparison of the measured and simulated (at 3.22\,GHz) fields when applying microwaves to each resonator in turn is shown in table\,\ref{bfieldstab}.

\begin{table}[htdp]
\caption{Measured microwave B-field amplitudes at the ion with 1\,mW at each input port of the trap (1.38mW input into the diplexer inputs), compared with the values from the HFSS simulation.  The simulation assumes \textbf{B}$_0$ is along $x$.}
\begin{center}
\begin{tabular}{|c|cc|cc|}
\hline
 & Measured &   & Simulated &  \\ 
\hline
&  & Angle to &  & Angle to \\ 
Port & $\mathbf{B}_{\mu w}$ ($\mu$T) & \textbf{B}$_0$ ($^\circ$) & $\mathbf{B}_{\mu w}$ ($\mu$T) & \textbf{B}$_0$ ($^\circ$)\\ 
\hline
1 & 12.9  & 62.0 & 20.1 & 64.7\\
2 & 24.4 & 3.0 & 20.0 & 1.4\\
3 & 12.5 & 61.4 & 20.1 & 65.0\\
\hline
\end{tabular}
\end{center}
\label{bfieldstab}
\end{table}%

\begin{figure}
\includegraphics[width=0.7\columnwidth]{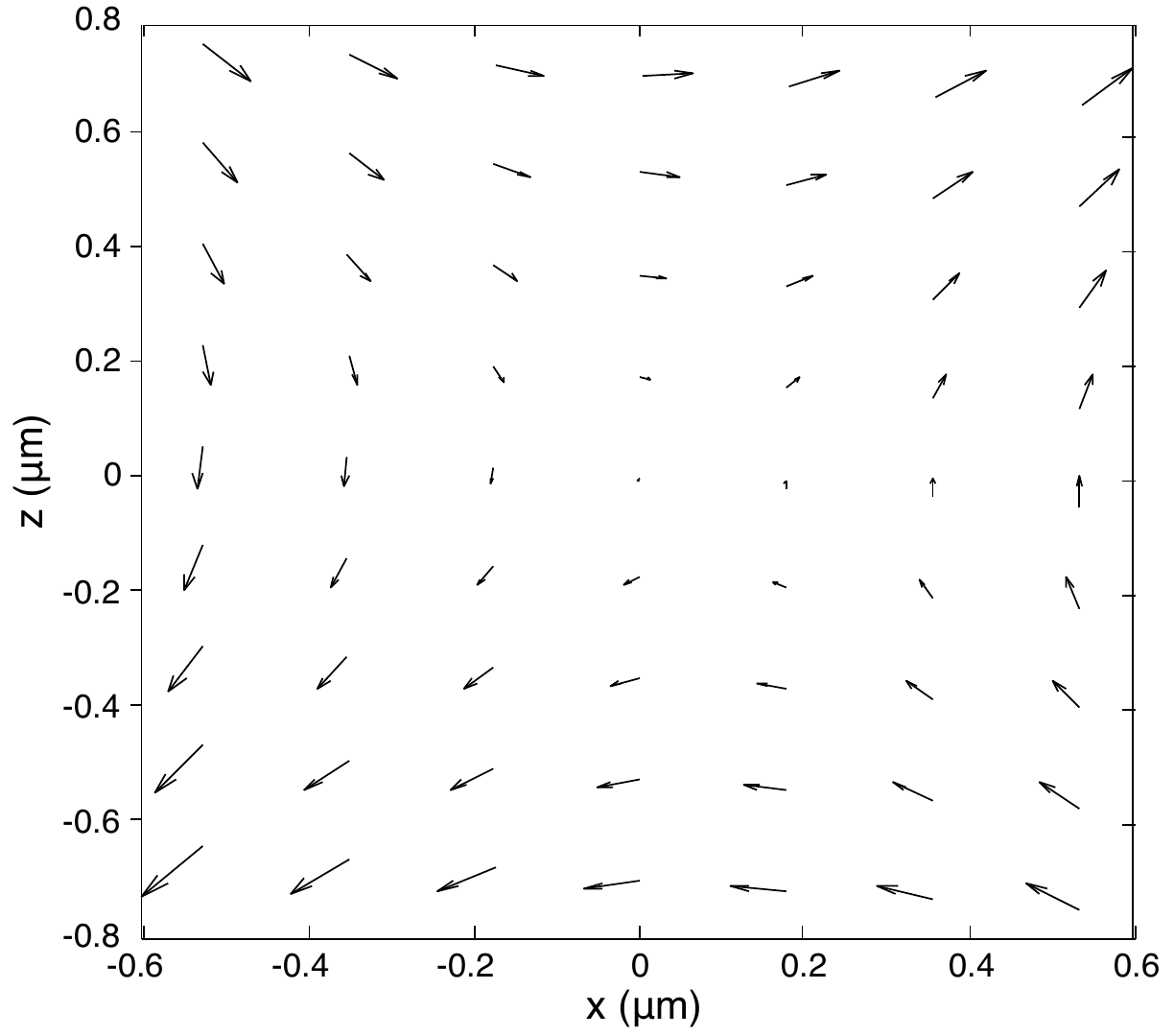}
\caption{\label{quad}Experimentally measured microwave B-field amplitude in the radial plane around the trap centre.  The field gradients at the trap centre are $\frac{\partial B_x}{\partial z}=0.304$\,T/m and $\frac{\partial B_z}{\partial x}=0.293$\,T/m.}
\end{figure}

In order to null the microwave field at the ion, the phase and amplitude of the microwave input  to each resonator was adjusted using IQ mixers\cite{Mine}.  Once nulling was achieved the trap potentials were adjusted to shift the ion to various points away from the trap's centre and at each of them the Rabi frequencies on the two transitions were measured (see fig.\,\ref{quad}).  The measured field gradients were $\frac{\partial B_x}{\partial z}=0.304$\,T/m and $\frac{\partial B_z}{\partial x}=0.293$\,T/m with 1.71\,mW, 0.48\,mW and 1.83\,mW applied to ports 1, 2 and 3 respectively.  The HFSS simulation predicts $\frac{\partial B_x}{\partial z}=\frac{\partial B_z}{\partial x}=0.289$\,T/m if the input powers in the simulation are adjusted such that the magnitude of the magnetic field from each electrode matches the experimental data.  This implies that the simulation of the field above the cavity matches experiment well but, as seen in the VNA data, there is some discrepancy in the coupling and resonant frequencies of the outer two resonators.  Based on thermal measurements on a similar neutral atom chip\cite{Gro04}, microwave input powers exceeding one Watt and gradients approaching 10\,T/m should be possible without damaging the trap.  In future experiments with this trap intend to utilise this gradient to implement multi-qubit entanglement.

%\subsection{Conclusions}

The integration of microwave circuitry demonstrated in this trap may be an important step towards a scalable ion trap quantum information processor.  In such a device the density of microwave circuitry could be increased by using a multi-layer semiconductor architecture, already demonstrated in other traps\cite{All12}.  Such a large scale device which would allow for many individually-addressed processing zones\cite{Mine}, a major simplification of the integration task compared to schemes where processing is carried out with lasers.  This trap design could also be developed towards a cryogenically-cooled experiment where a high-Q superconducting resonator could be combined with single photon microwave techniques to allow cavity QED experiments with charged particles in the microwave regime\cite{Schu11}.\\

%\begin{acknowledgments}
We would like to thank D. P. L. Aude Craik, L. Guidoni and B. Szymanski for technical assistance and D. Leibfried, C. Ospelkaus, A. M. Steane and U. Warring for useful discussions.  Microwave Office was generously donated to us by AWR Corporation.  This work was supported by an EPSRC Science \& Innovation Award and an EPSRC Visiting Scientist grant.\\
%\end{acknowledgments}

\bibliography{uwavetrap.bib}

\begin{thebibliography}{20}%
\makeatletter
\providecommand \@ifxundefined [1]{%
 \@ifx{#1\undefined}
}%
\providecommand \@ifnum [1]{%
 \ifnum #1\expandafter \@firstoftwo
 \else \expandafter \@secondoftwo
 \fi
}%
\providecommand \@ifx [1]{%
 \ifx #1\expandafter \@firstoftwo
 \else \expandafter \@secondoftwo
 \fi
}%
\providecommand \natexlab [1]{#1}%
\providecommand \enquote  [1]{``#1''}%
\providecommand \bibnamefont  [1]{#1}%
\providecommand \bibfnamefont [1]{#1}%
\providecommand \citenamefont [1]{#1}%
\providecommand \href@noop [0]{\@secondoftwo}%
\providecommand \href [0]{\begingroup \@sanitize@url \@href}%
\providecommand \@href[1]{\@@startlink{#1}\@@href}%
\providecommand \@@href[1]{\endgroup#1\@@endlink}%
\providecommand \@sanitize@url [0]{\catcode `\\12\catcode `\$12\catcode
  `\&12\catcode `\#12\catcode `\^12\catcode `\_12\catcode `\%12\relax}%
\providecommand \@@startlink[1]{}%
\providecommand \@@endlink[0]{}%
\providecommand \url  [0]{\begingroup\@sanitize@url \@url }%
\providecommand \@url [1]{\endgroup\@href {#1}{\urlprefix }}%
\providecommand \urlprefix  [0]{URL }%
\providecommand \Eprint [0]{\href }%
\providecommand \doibase [0]{http://dx.doi.org/}%
\providecommand \selectlanguage [0]{\@gobble}%
\providecommand \bibinfo  [0]{\@secondoftwo}%
\providecommand \bibfield  [0]{\@secondoftwo}%
\providecommand \translation [1]{[#1]}%
\providecommand \BibitemOpen [0]{}%
\providecommand \bibitemStop [0]{}%
\providecommand \bibitemNoStop [0]{.\EOS\space}%
\providecommand \EOS [0]{\spacefactor3000\relax}%
\providecommand \BibitemShut  [1]{\csname bibitem#1\endcsname}%
\let\auto@bib@innerbib\@empty
%</preamble>
\bibitem [{\citenamefont {Wineland}\ and\ \citenamefont
  {Leibfried}(2011)}]{Wineland11}%
  \BibitemOpen
  \bibfield  {author} {\bibinfo {author} {\bibfnamefont {D.~J.}\ \bibnamefont
  {Wineland}}\ and\ \bibinfo {author} {\bibfnamefont {D.}~\bibnamefont
  {Leibfried}},\ }\href@noop {} {\bibfield  {journal} {\bibinfo  {journal}
  {Laser Phys. Lett.}\ }\textbf {\bibinfo {volume} {8}},\ \bibinfo {pages}
  {175} (\bibinfo {year} {2011})}\BibitemShut {NoStop}%
\bibitem [{\citenamefont {Seidelin}\ \emph {et~al.}(2006)\citenamefont
  {Seidelin}, \citenamefont {Chiaverini}, \citenamefont {Reichle},
  \citenamefont {Bollinger}, \citenamefont {Leibfried}, \citenamefont
  {Britton}, \citenamefont {Wesenberg}, \citenamefont {Blakestad},
  \citenamefont {Epstein}, \citenamefont {Hume}, \citenamefont {Itano},
  \citenamefont {Jost}, \citenamefont {Langer}, \citenamefont {Ozeri},
  \citenamefont {Shiga},\ and\ \citenamefont {Wineland}}]{Sei06}%
  \BibitemOpen
  \bibfield  {author} {\bibinfo {author} {\bibfnamefont {S.}~\bibnamefont
  {Seidelin}}, \bibinfo {author} {\bibfnamefont {J.}~\bibnamefont
  {Chiaverini}}, \bibinfo {author} {\bibfnamefont {R.}~\bibnamefont {Reichle}},
  \bibinfo {author} {\bibfnamefont {J.~J.}\ \bibnamefont {Bollinger}}, \bibinfo
  {author} {\bibfnamefont {D.}~\bibnamefont {Leibfried}}, \bibinfo {author}
  {\bibfnamefont {J.}~\bibnamefont {Britton}}, \bibinfo {author} {\bibfnamefont
  {J.~H.}\ \bibnamefont {Wesenberg}}, \bibinfo {author} {\bibfnamefont {R.~B.}\
  \bibnamefont {Blakestad}}, \bibinfo {author} {\bibfnamefont {R.~J.}\
  \bibnamefont {Epstein}}, \bibinfo {author} {\bibfnamefont {D.~B.}\
  \bibnamefont {Hume}}, \bibinfo {author} {\bibfnamefont {W.~M.}\ \bibnamefont
  {Itano}}, \bibinfo {author} {\bibfnamefont {J.~D.}\ \bibnamefont {Jost}},
  \bibinfo {author} {\bibfnamefont {C.}~\bibnamefont {Langer}}, \bibinfo
  {author} {\bibfnamefont {R.}~\bibnamefont {Ozeri}}, \bibinfo {author}
  {\bibfnamefont {N.}~\bibnamefont {Shiga}}, \ and\ \bibinfo {author}
  {\bibfnamefont {D.~J.}\ \bibnamefont {Wineland}},\ }\href@noop {} {\bibfield
  {journal} {\bibinfo  {journal} {Phys. Rev. Lett.}\ }\textbf {\bibinfo
  {volume} {96}},\ \bibinfo {pages} {253003} (\bibinfo {year}
  {2006})}\BibitemShut {NoStop}%
\bibitem [{\citenamefont {Reichel}\ and\ \citenamefont
  {Vuletic}(2011)}]{AtomChips}%
  \BibitemOpen
  \bibinfo {editor} {\bibfnamefont {J.}~\bibnamefont {Reichel}}\ and\ \bibinfo
  {editor} {\bibfnamefont {V.}~\bibnamefont {Vuletic}},\ eds.,\ \href@noop {}
  {\emph {\bibinfo {title} {Atom Chips}}}\ (\bibinfo  {publisher} {Wiley VCH,
  Berlin},\ \bibinfo {year} {2011})\BibitemShut {NoStop}%
\bibitem [{\citenamefont {Ospelkaus}\ \emph {et~al.}(2011)\citenamefont
  {Ospelkaus}, \citenamefont {Warring}, \citenamefont {Colombe}, \citenamefont
  {Brown}, \citenamefont {Amini}, \citenamefont {Leibfried},\ and\
  \citenamefont {Wineland}}]{Osp11}%
  \BibitemOpen
  \bibfield  {author} {\bibinfo {author} {\bibfnamefont {C.}~\bibnamefont
  {Ospelkaus}}, \bibinfo {author} {\bibfnamefont {U.}~\bibnamefont {Warring}},
  \bibinfo {author} {\bibfnamefont {Y.}~\bibnamefont {Colombe}}, \bibinfo
  {author} {\bibfnamefont {K.~R.}\ \bibnamefont {Brown}}, \bibinfo {author}
  {\bibfnamefont {J.~M.}\ \bibnamefont {Amini}}, \bibinfo {author}
  {\bibfnamefont {D.}~\bibnamefont {Leibfried}}, \ and\ \bibinfo {author}
  {\bibfnamefont {D.~J.}\ \bibnamefont {Wineland}},\ }\href@noop {} {\bibfield
  {journal} {\bibinfo  {journal} {Nature (London)}\ }\textbf {\bibinfo {volume}
  {476}},\ \bibinfo {pages} {181} (\bibinfo {year} {2011})}\BibitemShut
  {NoStop}%
\bibitem [{\citenamefont {B\"ohi}\ \emph {et~al.}(2009)\citenamefont {B\"ohi},
  \citenamefont {Riedel}, \citenamefont {Hoffrogge}, \citenamefont {Reichel},
  \citenamefont {H\"ansch},\ and\ \citenamefont {Treutlein}}]{Boh09}%
  \BibitemOpen
  \bibfield  {author} {\bibinfo {author} {\bibfnamefont {P.}~\bibnamefont
  {B\"ohi}}, \bibinfo {author} {\bibfnamefont {M.~F.}\ \bibnamefont {Riedel}},
  \bibinfo {author} {\bibfnamefont {J.}~\bibnamefont {Hoffrogge}}, \bibinfo
  {author} {\bibfnamefont {J.}~\bibnamefont {Reichel}}, \bibinfo {author}
  {\bibfnamefont {T.~W.}\ \bibnamefont {H\"ansch}}, \ and\ \bibinfo {author}
  {\bibfnamefont {P.}~\bibnamefont {Treutlein}},\ }\href@noop {} {\bibfield
  {journal} {\bibinfo  {journal} {Nat. Phys.}\ }\textbf {\bibinfo {volume}
  {5}},\ \bibinfo {pages} {592} (\bibinfo {year} {2009})}\BibitemShut {NoStop}%
\bibitem [{\citenamefont {Brown}\ \emph {et~al.}(2011)\citenamefont {Brown},
  \citenamefont {Wilson}, \citenamefont {Colombe}, \citenamefont {Ospelkaus},
  \citenamefont {Meier}, \citenamefont {Knill}, \citenamefont {Leibfried},\
  and\ \citenamefont {Wineland}}]{Bro11}%
  \BibitemOpen
  \bibfield  {author} {\bibinfo {author} {\bibfnamefont {K.~R.}\ \bibnamefont
  {Brown}}, \bibinfo {author} {\bibfnamefont {A.~C.}\ \bibnamefont {Wilson}},
  \bibinfo {author} {\bibfnamefont {Y.}~\bibnamefont {Colombe}}, \bibinfo
  {author} {\bibfnamefont {C.}~\bibnamefont {Ospelkaus}}, \bibinfo {author}
  {\bibfnamefont {A.~M.}\ \bibnamefont {Meier}}, \bibinfo {author}
  {\bibfnamefont {E.}~\bibnamefont {Knill}}, \bibinfo {author} {\bibfnamefont
  {D.}~\bibnamefont {Leibfried}}, \ and\ \bibinfo {author} {\bibfnamefont
  {D.~J.}\ \bibnamefont {Wineland}},\ }\href@noop {} {\bibfield  {journal}
  {\bibinfo  {journal} {Phys. Rev. A}\ }\textbf {\bibinfo {volume} {84}},\
  \bibinfo {pages} {030303} (\bibinfo {year} {2011})}\BibitemShut {NoStop}%
\bibitem [{\citenamefont {Antohi}(2011)}]{Ant11}%
  \BibitemOpen
  \bibfield  {author} {\bibinfo {author} {\bibfnamefont {P.}~\bibnamefont
  {Antohi}},\ }\href@noop {} {Ph.D. thesis},\ \bibinfo  {school} {Massachusetts
  Institute of Technology} (\bibinfo {year} {2011})\BibitemShut {NoStop}%
\bibitem [{\citenamefont {Lucas}\ \emph {et~al.}()\citenamefont {Lucas},
  \citenamefont {Keitch}, \citenamefont {Home}, \citenamefont {Imreh},
  \citenamefont {McDonnell}, \citenamefont {Stacey}, \citenamefont {Szwer},\
  and\ \citenamefont {Steane}}]{Lucas2007}%
  \BibitemOpen
  \bibfield  {author} {\bibinfo {author} {\bibfnamefont {D.~M.}\ \bibnamefont
  {Lucas}}, \bibinfo {author} {\bibfnamefont {B.~C.}\ \bibnamefont {Keitch}},
  \bibinfo {author} {\bibfnamefont {J.~P.}\ \bibnamefont {Home}}, \bibinfo
  {author} {\bibfnamefont {G.}~\bibnamefont {Imreh}}, \bibinfo {author}
  {\bibfnamefont {M.~J.}\ \bibnamefont {McDonnell}}, \bibinfo {author}
  {\bibfnamefont {D.~N.}\ \bibnamefont {Stacey}}, \bibinfo {author}
  {\bibfnamefont {D.~J.}\ \bibnamefont {Szwer}}, \ and\ \bibinfo {author}
  {\bibfnamefont {A.~M.}\ \bibnamefont {Steane}},\ }\href@noop {} {\bibinfo
  {journal} {arXiv:quant-ph/0710.4421v1 (unpublished)}\ }\BibitemShut {NoStop}%
\bibitem [{\citenamefont {Ospelkaus}\ \emph {et~al.}(2008)\citenamefont
  {Ospelkaus}, \citenamefont {Langer}, \citenamefont {Amini}, \citenamefont
  {Brown}, \citenamefont {Leibfried},\ and\ \citenamefont {Wineland}}]{Osp08}%
  \BibitemOpen
\bibfield  {journal} {  }\bibfield  {author} {\bibinfo {author} {\bibfnamefont
  {C.}~\bibnamefont {Ospelkaus}}, \bibinfo {author} {\bibfnamefont {C.~E.}\
  \bibnamefont {Langer}}, \bibinfo {author} {\bibfnamefont {J.~M.}\
  \bibnamefont {Amini}}, \bibinfo {author} {\bibfnamefont {K.~R.}\ \bibnamefont
  {Brown}}, \bibinfo {author} {\bibfnamefont {D.}~\bibnamefont {Leibfried}}, \
  and\ \bibinfo {author} {\bibfnamefont {D.~J.}\ \bibnamefont {Wineland}},\
  }\href@noop {} {\bibfield  {journal} {\bibinfo  {journal} {Phys. Rev. Lett.}\
  }\textbf {\bibinfo {volume} {101}},\ \bibinfo {pages} {090502} (\bibinfo
  {year} {2008})}\BibitemShut {NoStop}%
\bibitem [{\citenamefont {Pozar}(2005)}]{Poz05}%
  \BibitemOpen
  \bibfield  {author} {\bibinfo {author} {\bibfnamefont {D.~M.}\ \bibnamefont
  {Pozar}},\ }\href@noop {} {\emph {\bibinfo {title} {Microwave
  Engineering}}},\ \bibinfo {edition} {3rd}\ ed.\ (\bibinfo  {publisher}
  {Wiley, New York},\ \bibinfo {year} {2005})\BibitemShut {NoStop}%
\bibitem [{\citenamefont {Zhu}, \citenamefont {Shi},\ and\ \citenamefont
  {Menzel}(2005)}]{Zhu05}%
  \BibitemOpen
  \bibfield  {author} {\bibinfo {author} {\bibfnamefont {L.}~\bibnamefont
  {Zhu}}, \bibinfo {author} {\bibfnamefont {H.}~\bibnamefont {Shi}}, \ and\
  \bibinfo {author} {\bibfnamefont {W.}~\bibnamefont {Menzel}},\ }\href@noop {}
  {\bibfield  {journal} {\bibinfo  {journal} {IEEE Microw. Wireless Compon.
  Lett.}\ }\textbf {\bibinfo {volume} {15}},\ \bibinfo {pages} {13} (\bibinfo
  {year} {2005})}\BibitemShut {NoStop}%
\bibitem [{\citenamefont {Allcock}\ \emph {et~al.}(2010)\citenamefont
  {Allcock}, \citenamefont {Sherman}, \citenamefont {Stacey}, \citenamefont
  {Burrell}, \citenamefont {Curtis}, \citenamefont {Imreh}, \citenamefont
  {Linke}, \citenamefont {Szwer}, \citenamefont {Webster}, \citenamefont
  {Steane},\ and\ \citenamefont {Lucas}}]{Allcock10}%
  \BibitemOpen
  \bibfield  {author} {\bibinfo {author} {\bibfnamefont {D.~T.~C.}\
  \bibnamefont {Allcock}}, \bibinfo {author} {\bibfnamefont {J.~A.}\
  \bibnamefont {Sherman}}, \bibinfo {author} {\bibfnamefont {D.~N.}\
  \bibnamefont {Stacey}}, \bibinfo {author} {\bibfnamefont {A.~H.}\
  \bibnamefont {Burrell}}, \bibinfo {author} {\bibfnamefont {M.~J.}\
  \bibnamefont {Curtis}}, \bibinfo {author} {\bibfnamefont {G.}~\bibnamefont
  {Imreh}}, \bibinfo {author} {\bibfnamefont {N.~M.}\ \bibnamefont {Linke}},
  \bibinfo {author} {\bibfnamefont {D.~J.}\ \bibnamefont {Szwer}}, \bibinfo
  {author} {\bibfnamefont {S.~C.}\ \bibnamefont {Webster}}, \bibinfo {author}
  {\bibfnamefont {A.~M.}\ \bibnamefont {Steane}}, \ and\ \bibinfo {author}
  {\bibfnamefont {D.~M.}\ \bibnamefont {Lucas}},\ }\href@noop {} {\bibfield
  {journal} {\bibinfo  {journal} {New J. Phys.}\ }\textbf {\bibinfo {volume}
  {12}},\ \bibinfo {pages} {053026} (\bibinfo {year} {2010})}\BibitemShut
  {NoStop}%
\bibitem [{\citenamefont {Allcock}(2011)}]{Mine}%
  \BibitemOpen
  \bibfield  {author} {\bibinfo {author} {\bibfnamefont {D.~T.~C.}\
  \bibnamefont {Allcock}},\ }\href@noop {} {Ph.D. thesis},\ \bibinfo  {school}
  {University of Oxford} (\bibinfo {year} {2011})\BibitemShut {NoStop}%
\bibitem [{\citenamefont {Lucas}\ \emph {et~al.}(2004)\citenamefont {Lucas},
  \citenamefont {Ramos}, \citenamefont {Home}, \citenamefont {McDonnell},
  \citenamefont {Nakayama}, \citenamefont {Stacey}, \citenamefont {Webster},
  \citenamefont {Stacey},\ and\ \citenamefont {Steane}}]{Lucas04}%
  \BibitemOpen
  \bibfield  {author} {\bibinfo {author} {\bibfnamefont {D.~M.}\ \bibnamefont
  {Lucas}}, \bibinfo {author} {\bibfnamefont {A.}~\bibnamefont {Ramos}},
  \bibinfo {author} {\bibfnamefont {J.~P.}\ \bibnamefont {Home}}, \bibinfo
  {author} {\bibfnamefont {M.~J.}\ \bibnamefont {McDonnell}}, \bibinfo {author}
  {\bibfnamefont {S.}~\bibnamefont {Nakayama}}, \bibinfo {author}
  {\bibfnamefont {J.~P.}\ \bibnamefont {Stacey}}, \bibinfo {author}
  {\bibfnamefont {S.~C.}\ \bibnamefont {Webster}}, \bibinfo {author}
  {\bibfnamefont {D.~N.}\ \bibnamefont {Stacey}}, \ and\ \bibinfo {author}
  {\bibfnamefont {A.~M.}\ \bibnamefont {Steane}},\ }\href@noop {} {\bibfield
  {journal} {\bibinfo  {journal} {Phys. Rev. A}\ }\textbf {\bibinfo {volume}
  {69}},\ \bibinfo {pages} {012711} (\bibinfo {year} {2004})}\BibitemShut
  {NoStop}%
\bibitem [{\citenamefont {Wesenberg}\ \emph {et~al.}(2007)\citenamefont
  {Wesenberg}, \citenamefont {Epstein}, \citenamefont {Leibfried},
  \citenamefont {Blakestad}, \citenamefont {Britton}, \citenamefont {Home},
  \citenamefont {Itano}, \citenamefont {Jost}, \citenamefont {Knill},
  \citenamefont {Langer}, \citenamefont {Ozeri}, \citenamefont {Seidelin},\
  and\ \citenamefont {Wineland}}]{Wes07}%
  \BibitemOpen
  \bibfield  {author} {\bibinfo {author} {\bibfnamefont {J.~H.}\ \bibnamefont
  {Wesenberg}}, \bibinfo {author} {\bibfnamefont {R.~J.}\ \bibnamefont
  {Epstein}}, \bibinfo {author} {\bibfnamefont {D.}~\bibnamefont {Leibfried}},
  \bibinfo {author} {\bibfnamefont {R.~B.}\ \bibnamefont {Blakestad}}, \bibinfo
  {author} {\bibfnamefont {J.}~\bibnamefont {Britton}}, \bibinfo {author}
  {\bibfnamefont {J.~P.}\ \bibnamefont {Home}}, \bibinfo {author}
  {\bibfnamefont {W.~M.}\ \bibnamefont {Itano}}, \bibinfo {author}
  {\bibfnamefont {J.~D.}\ \bibnamefont {Jost}}, \bibinfo {author}
  {\bibfnamefont {E.}~\bibnamefont {Knill}}, \bibinfo {author} {\bibfnamefont
  {C.}~\bibnamefont {Langer}}, \bibinfo {author} {\bibfnamefont
  {R.}~\bibnamefont {Ozeri}}, \bibinfo {author} {\bibfnamefont
  {S.}~\bibnamefont {Seidelin}}, \ and\ \bibinfo {author} {\bibfnamefont
  {D.~J.}\ \bibnamefont {Wineland}},\ }\href@noop {} {\bibfield  {journal}
  {\bibinfo  {journal} {Phys. Rev. A}\ }\textbf {\bibinfo {volume} {76}},\
  \bibinfo {pages} {053416} (\bibinfo {year} {2007})}\BibitemShut {NoStop}%
\bibitem [{\citenamefont {Hite}\ \emph {et~al.}(2012)\citenamefont {Hite},
  \citenamefont {Colombe}, \citenamefont {Wilson}, \citenamefont {Brown},
  \citenamefont {Warring}, \citenamefont {J{\"o}rdens}, \citenamefont {Jost},
  \citenamefont {McKay}, \citenamefont {Pappas}, \citenamefont {Leibfried},\
  and\ \citenamefont {Wineland}}]{Hite12}%
  \BibitemOpen
  \bibfield  {author} {\bibinfo {author} {\bibfnamefont {D.~A.}\ \bibnamefont
  {Hite}}, \bibinfo {author} {\bibfnamefont {Y.}~\bibnamefont {Colombe}},
  \bibinfo {author} {\bibfnamefont {A.~C.}\ \bibnamefont {Wilson}}, \bibinfo
  {author} {\bibfnamefont {K.~R.}\ \bibnamefont {Brown}}, \bibinfo {author}
  {\bibfnamefont {U.}~\bibnamefont {Warring}}, \bibinfo {author} {\bibfnamefont
  {R.}~\bibnamefont {J{\"o}rdens}}, \bibinfo {author} {\bibfnamefont {J.~D.}\
  \bibnamefont {Jost}}, \bibinfo {author} {\bibfnamefont {K.}~\bibnamefont
  {McKay}}, \bibinfo {author} {\bibfnamefont {D.~P.}\ \bibnamefont {Pappas}},
  \bibinfo {author} {\bibfnamefont {D.}~\bibnamefont {Leibfried}}, \ and\
  \bibinfo {author} {\bibfnamefont {D.~J.}\ \bibnamefont {Wineland}},\
  }\href@noop {} {\bibfield  {journal} {\bibinfo  {journal} {Phys. Rev. Lett.}\
  }\textbf {\bibinfo {volume} {109}},\ \bibinfo {pages} {103001} (\bibinfo
  {year} {2012})}\BibitemShut {NoStop}%
\bibitem [{\citenamefont {Berkeland}\ \emph {et~al.}(1998)\citenamefont
  {Berkeland}, \citenamefont {Miller}, \citenamefont {Bergquist}, \citenamefont
  {Itano},\ and\ \citenamefont {Wineland}}]{Ber98}%
  \BibitemOpen
  \bibfield  {author} {\bibinfo {author} {\bibfnamefont {D.~J.}\ \bibnamefont
  {Berkeland}}, \bibinfo {author} {\bibfnamefont {J.~D.}\ \bibnamefont
  {Miller}}, \bibinfo {author} {\bibfnamefont {J.~C.}\ \bibnamefont
  {Bergquist}}, \bibinfo {author} {\bibfnamefont {W.~M.}\ \bibnamefont
  {Itano}}, \ and\ \bibinfo {author} {\bibfnamefont {D.~J.}\ \bibnamefont
  {Wineland}},\ }\href@noop {} {\bibfield  {journal} {\bibinfo  {journal} {J.
  App. Phys.}\ }\textbf {\bibinfo {volume} {83}},\ \bibinfo {pages} {5025}
  (\bibinfo {year} {1998})}\BibitemShut {NoStop}%
\bibitem [{\citenamefont {Groth}\ \emph {et~al.}(2004)\citenamefont {Groth},
  \citenamefont {Kr\"{u}ger}, \citenamefont {Wildermuth}, \citenamefont
  {Folman}, \citenamefont {Fernholz}, \citenamefont {Mahalu}, \citenamefont
  {Bar-{J}oseph},\ and\ \citenamefont {Schmiedmayer}}]{Gro04}%
  \BibitemOpen
  \bibfield  {author} {\bibinfo {author} {\bibfnamefont {S.}~\bibnamefont
  {Groth}}, \bibinfo {author} {\bibfnamefont {P.}~\bibnamefont {Kr\"{u}ger}},
  \bibinfo {author} {\bibfnamefont {S.}~\bibnamefont {Wildermuth}}, \bibinfo
  {author} {\bibfnamefont {R.}~\bibnamefont {Folman}}, \bibinfo {author}
  {\bibfnamefont {T.}~\bibnamefont {Fernholz}}, \bibinfo {author}
  {\bibfnamefont {D.}~\bibnamefont {Mahalu}}, \bibinfo {author} {\bibfnamefont
  {I.}~\bibnamefont {Bar-{J}oseph}}, \ and\ \bibinfo {author} {\bibfnamefont
  {J.}~\bibnamefont {Schmiedmayer}},\ }\href@noop {} {\bibfield  {journal}
  {\bibinfo  {journal} {Appl. Phys. Lett.}\ }\textbf {\bibinfo {volume} {85}},\
  \bibinfo {pages} {2980} (\bibinfo {year} {2004})}\BibitemShut {NoStop}%
\bibitem [{\citenamefont {Allcock}\ \emph {et~al.}(2012)\citenamefont
  {Allcock}, \citenamefont {Harty}, \citenamefont {Janacek}, \citenamefont
  {Linke}, \citenamefont {Ballance}, \citenamefont {Steane}, \citenamefont
  {Lucas}, \citenamefont {{Jarecki Jr}.}, \citenamefont {Habermehl},
  \citenamefont {Blain}, \citenamefont {Stick},\ and\ \citenamefont
  {Moehring}}]{All12}%
  \BibitemOpen
  \bibfield  {author} {\bibinfo {author} {\bibfnamefont {D.~T.~C.}\
  \bibnamefont {Allcock}}, \bibinfo {author} {\bibfnamefont {T.~P.}\
  \bibnamefont {Harty}}, \bibinfo {author} {\bibfnamefont {H.~A.}\ \bibnamefont
  {Janacek}}, \bibinfo {author} {\bibfnamefont {N.~M.}\ \bibnamefont {Linke}},
  \bibinfo {author} {\bibfnamefont {C.~J.}\ \bibnamefont {Ballance}}, \bibinfo
  {author} {\bibfnamefont {A.~M.}\ \bibnamefont {Steane}}, \bibinfo {author}
  {\bibfnamefont {D.~M.}\ \bibnamefont {Lucas}}, \bibinfo {author}
  {\bibfnamefont {R.~L.}\ \bibnamefont {{Jarecki Jr}.}}, \bibinfo {author}
  {\bibfnamefont {S.~D.}\ \bibnamefont {Habermehl}}, \bibinfo {author}
  {\bibfnamefont {M.~G.}\ \bibnamefont {Blain}}, \bibinfo {author}
  {\bibfnamefont {D.}~\bibnamefont {Stick}}, \ and\ \bibinfo {author}
  {\bibfnamefont {D.~L.}\ \bibnamefont {Moehring}},\ }\href@noop {} {\bibfield
  {journal} {\bibinfo  {journal} {Appl. Phys. B: Lasers Opt.}\ }\textbf
  {\bibinfo {volume} {107}},\ \bibinfo {pages} {913} (\bibinfo {year}
  {2012})}\BibitemShut {NoStop}%
\bibitem [{\citenamefont {Schuster}\ \emph {et~al.}(2011)\citenamefont
  {Schuster}, \citenamefont {Bishop}, \citenamefont {Chuang}, \citenamefont
  {DeMille},\ and\ \citenamefont {Schoelkopf}}]{Schu11}%
  \BibitemOpen
  \bibfield  {author} {\bibinfo {author} {\bibfnamefont {D.~I.}\ \bibnamefont
  {Schuster}}, \bibinfo {author} {\bibfnamefont {L.~S.}\ \bibnamefont
  {Bishop}}, \bibinfo {author} {\bibfnamefont {I.~L.}\ \bibnamefont {Chuang}},
  \bibinfo {author} {\bibfnamefont {D.}~\bibnamefont {DeMille}}, \ and\
  \bibinfo {author} {\bibfnamefont {R.~J.}\ \bibnamefont {Schoelkopf}},\
  }\href@noop {} {\bibfield  {journal} {\bibinfo  {journal} {Phys. Rev. A}\
  }\textbf {\bibinfo {volume} {83}},\ \bibinfo {pages} {012311} (\bibinfo
  {year} {2011})}\BibitemShut {NoStop}%
\end{thebibliography}%

\end{document}